\documentclass[12pt, referee]{aastex}
\usepackage[dvips]{color}
\usepackage{natbib}
\usepackage{lineno}

\def\gtsima{$\;\buildrel > \over \sim \;$}
\def\simgt{\lower.5ex \hbox{\gtsima}}
\def\ltsima{$\;\buildrel < \over \sim \;$}
\def\simlt{\lower.5ex \hbox{\ltsima}}
\def \CHp{\ifmmode{\rm CH^+}\else{$\rm CH^+$}\fi}
\def \HH{\ifmmode{\rm H_2}\else{$\rm H_2$}\fi}
\def \Cp{\ifmmode{\rm C^+}\else{$\rm C^+$}\fi} 
\def \cc    {\ifmmode{\,{\rm cm}^{-3}}\else{$\,{\rm cm}^{-3}$}\fi}
\def \dens{\ifmmode{n_{\rm H}}\else{$n_{\rm H}$}\fi}
\def \kms   {\ifmmode{\,{\rm km}\,{\rm s}^{-1}}\else{km s$^{-1}$}\fi} 

\begin{document}

\title{Infrared line emissions from atoms and atomic ions in NGC 7027: improved wavelength determinations for infrared metal lines and a probable detection of Zn$^{5+}$}

\author{David A. Neufeld\altaffilmark{1}}

\altaffiltext{1}{William H.\ Miller Department of Physics \& Astronomy, Johns Hopkins University, Baltimore, MD 21218, USA}

\begin{abstract}

An infrared L- and M-band spectral survey, performed toward the young planetary nebula 
NGC 7027 with the iSHELL instrument on NASA's Infrared Telescope Facility (IRTF), 
has revealed more than 20 vibrational lines of the molecules HeH$^+$, H$_2$, 
and CH$^+$ and more than 50 spectral lines of atoms and atomic ions.  
The present paper focuses on the atomic line emissions, the molecular lines 
having been discussed in two previous publications (Neufeld et al.\ 2020, 2021).   
The atomic lines detected with high confidence in the 2.951 - 5.24 micron region covered 
(incompletely) by this survey comprise (1) six collisionally-excited lines 
of  metal ions that had previously been 
identified in astrophysical nebulae but for which the  
present observations provide the most accurate wavelength determinations 
obtained to date; (2) a spectral line at 4.6895$\rm \,\mu$m, not previously 
reported, for which the probable identification is the $^4F_{7/2} - $~$^4F_{9/2}$ fine structure  transition of [Zn~VI]; (3) 39 recombination lines of H and He$^+$, with upper states of principal quantum number up to 38 (H) or 24 (He$^+$); (4) 10 recombination lines of the multielectron species He, C$^{2+}$, and C$^{3+}$.

\end{abstract}

\keywords{Planetary nebulae (1249), Interstellar line emission (844), Infrared astronomy (786), Atomic spectroscopy (2099)}

\section{Introduction}

Planetary nebulae are heated, photoionized and photodissociated by the continuum emission from a hot central star with an effective temperature that may exceed 200,000~K.  They are characterized by line emission from a multitude of species that include neutral atoms, atomic ions (some -- such as Ne$^{4+}$ -- with appearance potentials in excess of 100 eV), diatomic and polyatomic molecules, and molecular ions.
The latter include the HeH$^+$ cation, the astrophysical discovery of which was
reported recently toward the young, carbon-rich planetary 
nebula NGC 7027 (G\"usten et al.\ 2019). 
This detection of HeH$^+$, obtained through SOFIA/GREAT observations of 
its fundamental rotational 
transition in the far-IR spectral region, was followed by the detection of two
HeH$^+$ rovibrational lines using the infrared spectrograph, iSHELL, on the IRTF on Maunakea (Neufeld et al.\ 2020, hereafter Paper I).  The iSHELL observations reported in 
Paper I also led to the serendipitous discovery of rovibrational emissions from the CH$^+$ molecular ion, motivating a more extensive L- and M-band survey discussed in Neufeld et al.\ (2021; hereafter Paper II).  

While the previous papers in this series have focused on the infrared molecular emissions observed from NGC 7027, the present paper discusses the many spectral lines of atoms and atomic ions that have also been detected.  A recent discussion of the source has been presented in Papers I and II and will not repeated here.  In Section 2 below, I discuss the iSHELL observations performed toward NGC 7027 and the methods used to reduce the resulting data.  The results are presented in Section 3, and the key findings are discussed in Section 4: these include the most accurate wavelength determinations yet obtained for several fine structure transitions; the likely discovery of a fine structure line of [Zn~VI]; and a comparison of the relative strengths of the observed hydrogen and helium recombination lines with theoretical predictions.  A summary follows in Section 5.

\section{Observations and data reduction}

The observations reported in Paper I were performed at the IRTF on 2019 July 19 and September 6th UT with the iSHELL spectrograph (Rayner et al. 2016) in its {\tt Lp1}
and {\tt Lp2} modes respectively.  These were followed on 2020 July 8, 13, 14, and 
15 July UT by observations with the {\tt Lp3}, {\tt L2}, {\tt L1}, {\tt M1}, and {\tt M2} settings. Unfortunately, the night of 15 July UT was mainly cloudy and no usable data could be acquired with the {\tt M2} setting.  Because the free spectral range of the spectrograph cannot be covered fully by the M-band detector, two settings ({\tt M1} and {\tt M2}) are needed to obtain a full M-band spectrum; without usable data from the 
{\tt M2} setting, the M-band spectrum obtained from these observations therefore shows small wavelength gaps between the spectrograph orders.  These gaps 
account for roughly 8$\%$ percent of the M-band (4.517 -- 5.239 $\rm \,\mu$m) 
spectral coverage.  Moreover, in the region of spectral overlap between the {\tt L1} and 
{\tt L2} settings, the ${\tt L2}$ observations performed on 2020 July 13 yielded data of considerably higher quality than the {\tt L1} observations; moreover, the spectral region shortward of 2.951 $\rm \,\mu$m
that is covered by {\tt L1} but not {\tt L2} was severely affected by telluric lines.  The 
{\tt L1} data obtained on 2020 July 14 were therefore of little utility and were
also excluded from the analysis.  The resultant survey therefore provides complete 
coverage of the L-band spectral region from 2.951 - 4.155 $\rm \,\mu$m.  

The observing  
and data reduction procedures were similar for all observations 
and were described in detail in Paper I {and summarized in Paper II.  The
observations involved the
use of a narrow slit -- of width 0\farcs375 and length $15^{\prime\prime}$ -- that was 
oriented along the minor axis of the nebula at position angle 59\degr$\,$ East of North.
The spectral resolving power achieved with this 0\farcs375 slit is $\lambda/\Delta \lambda = 80,000.$ The wavelength calibration procedure made use of telluric lines (along with lines from a Thorium-Argon arc lamp in the case of the {\tt L1} mode); the accuracy of the  wavelength measurements it enabled for the spectral lines observed from NGC 7027 is discussed in Section 4.1 below.}  


\section{Results}

\begin{figure}
\includegraphics[scale=0.7]{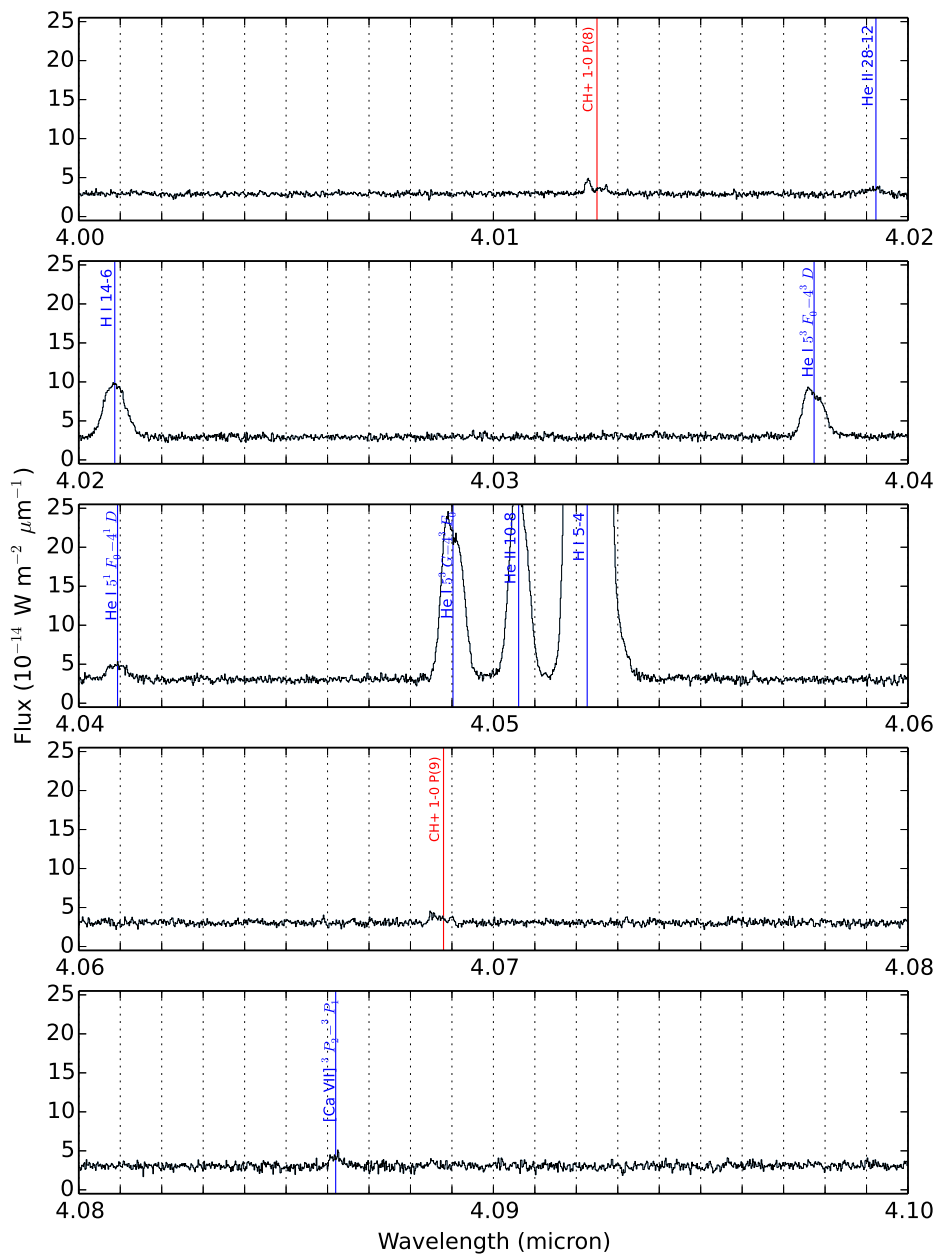}
\caption{Sample spectrum covering the $4.0 - 4.1 \,\mu$m range, summed over
the full slit length.  The wavelength scale is for an assumed LSR velocity of
25$\,\rm km\,s^{-1}$ for the source.  {This value is the mean LSR velocity of the ionized nebula as determined (Paper I) from a Gaussian fit to the H \small{I} 19 -- 6 line.}} 
\end{figure}

Figure 1, covering the 4.0 - 4.1 $\rm \,\mu$m region, shows an example of the full-slit spectra that are obtained by summing the signal over all 15 extraction regions.   Ten emission lines are readily identified in this small fraction ($\sim 6$ percent) of the available spectral coverage.  For the same example spectral region, Figure 2 provides information about the spatial variation of the intensity along the slit.  Here, the vertical axis represents the distance along the slit, and the intensity is color-coded on a logarithmic scale as a function of position and wavelength.  For an isolated spectral line, Figure 2 shows a position-velocity ($PV$) diagram of the type presented in Paper II (their Figures 2 -- 4).  The ring-like morphology typically exhibited here is precisely that expected for an expanding shell when long-slit spectroscopy is performed with the slit positioned along the diameter.

\begin{figure}
\includegraphics[scale=0.7]{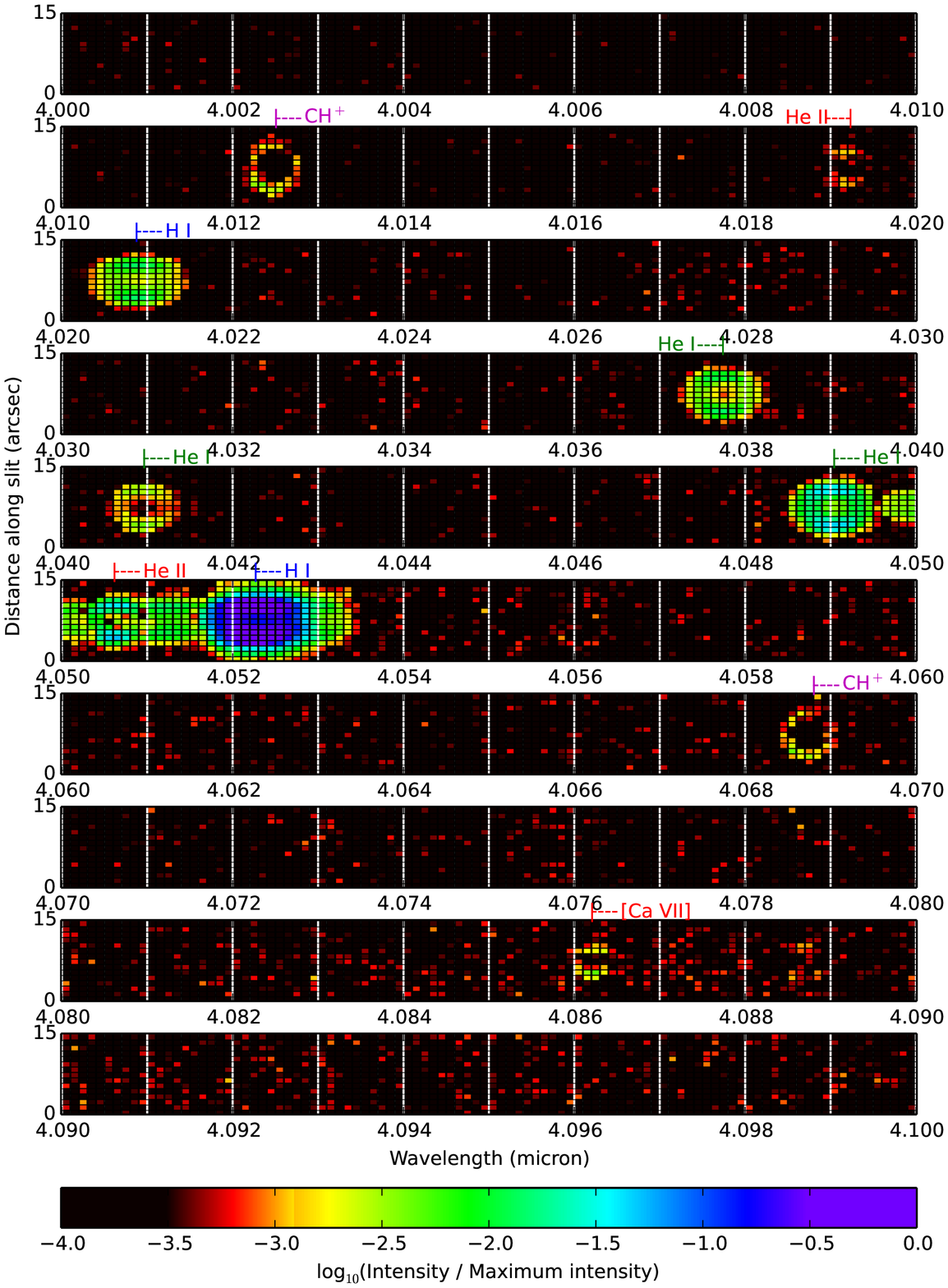}
\caption{Sample Position-Wavelength diagram covering the $4.0 - 4.1 \,\mu$m range,
showing the variation of the intensity along the slit.  The color bar shows -- on a logarithmic scale -- the intensity relative to the maximum intensity achieved within the $4.0 - 4.1 \,\mu$m wavelength range.
} 
\end{figure}

\begin{figure}
\includegraphics[scale=0.7]{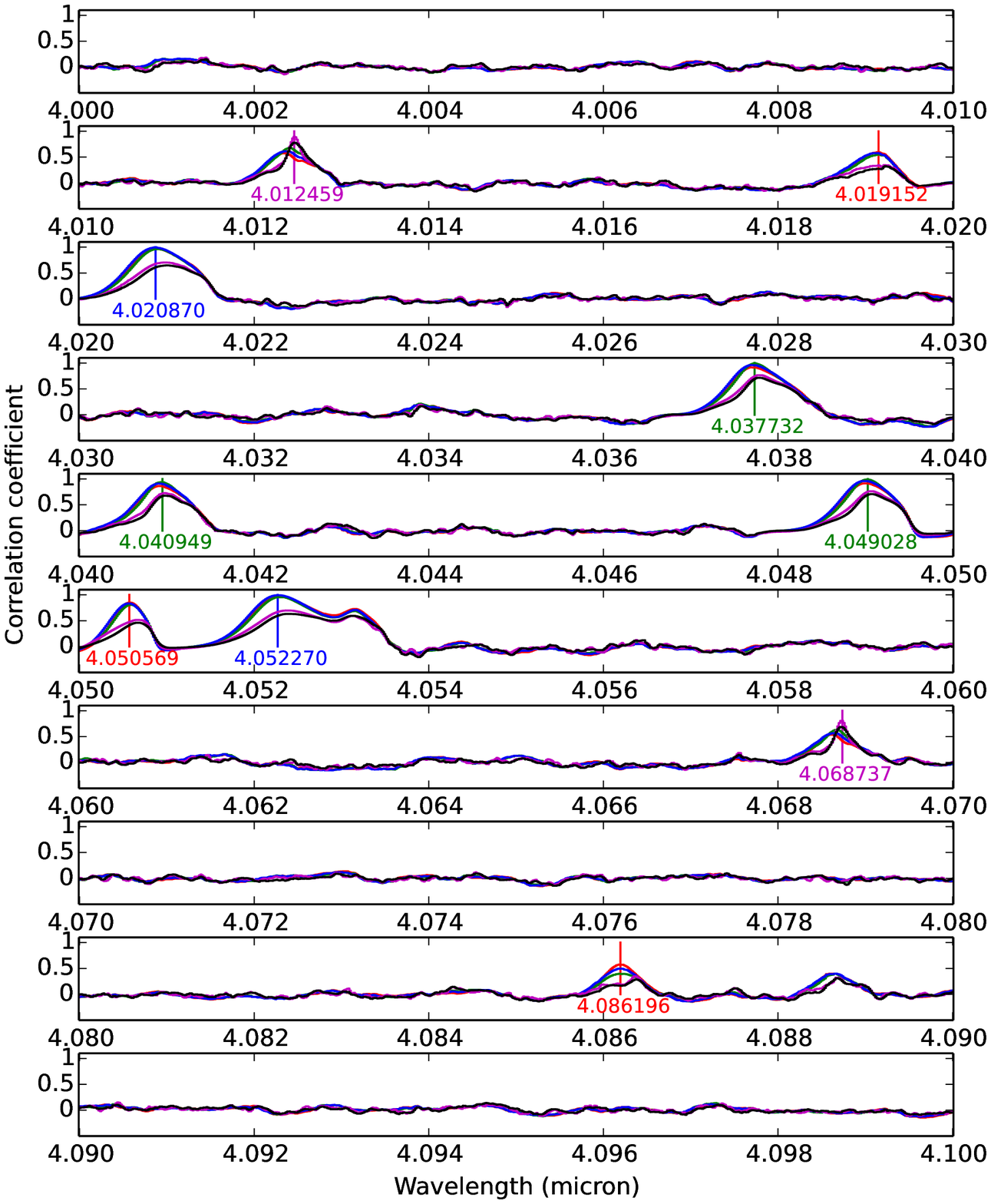}
\caption{Sample correlation coefficient plot, covering the $4.0 - 4.1 \,\mu$m range,
for five template transitions: He\small{ II} 7 -- 6 (red), He\small{ I} $5^3F^0-4^3D$ (green), 
H\small{ I} 14 -- 6 (blue), CH$^+ \, v=1-0\,P(5)$ (magenta), H$_2 \, v=1-0\,O(5)$ (black)}

\end{figure}

To implement a uniform criterion for line detection, I have developed a template-fitting  algorithm.  Here, the $PV$-diagrams for five spectral lines are adopted as templates.  The template transitions are all unblended, detected at a high signal-noise ratio, and have rest wavelengths that are known to high accuracy.  They are the He\small{ II} 7 -- 6, He\small{ I} $5^3F^0-4^3D$, H\small{ I} $14 - 6$, CH$^+ \, v=1-0\,P(5)$, and H$_2 \, v=1-0\,O(5)$ transitions, each of which shows a slightly different morphology in the $PV$-diagram.  These templates are then shifted in wavelength
and the linear correlation coefficient between these templates and the observed position-wavelength diagram (e.g. Figure 2) is computed as a function of wavelength.  Figure 3 shows the resultant correlation coefficients, plotted in red, green blue, magenta and black respectively for the five transitions listed above.  Local maxima
appear at the rest wavelengths of possible lines that are detected.  Moreover, a possible line identification may be suggested by observing which of the five templates yields the largest correlation coefficient.  Vertical lines in Figure 3 indicate the locations of the local maxima and are labeled with the rest wavelength in micron.  
These lines and labels are color-coded with the same colors as the plotted curves.  For all nine of the
spectral lines attributable to species in the template set (i.e.\ to H, He, He$^+$, H$_2$ or CH$^+$), this methodology yields the correct line identification, revealing two lines of CH$^+$, three lines of He~{\small I}, two lines of He {\small II}, and two lines of 
H {\small I}.  The 10th line shown in Figure 3, at a wavelength of 4.0862$\,\mu$m, is a fine structure transition of [Ca VII]\footnote{A secondary local maximum appearing slightly redward of the strong feature at $4.0523\, \rm \mu m$ (H \small{I} $5-4$; Brackett-$\alpha$) lacks any obvious identification as a distinct
line and may originate from redshifted ionized gas; it is also apparent as a red wing on the Brackett-$\alpha$ line in the full-slit spectra plotted in Figure 1.}.  
\begin{figure}
\includegraphics[scale=0.7]{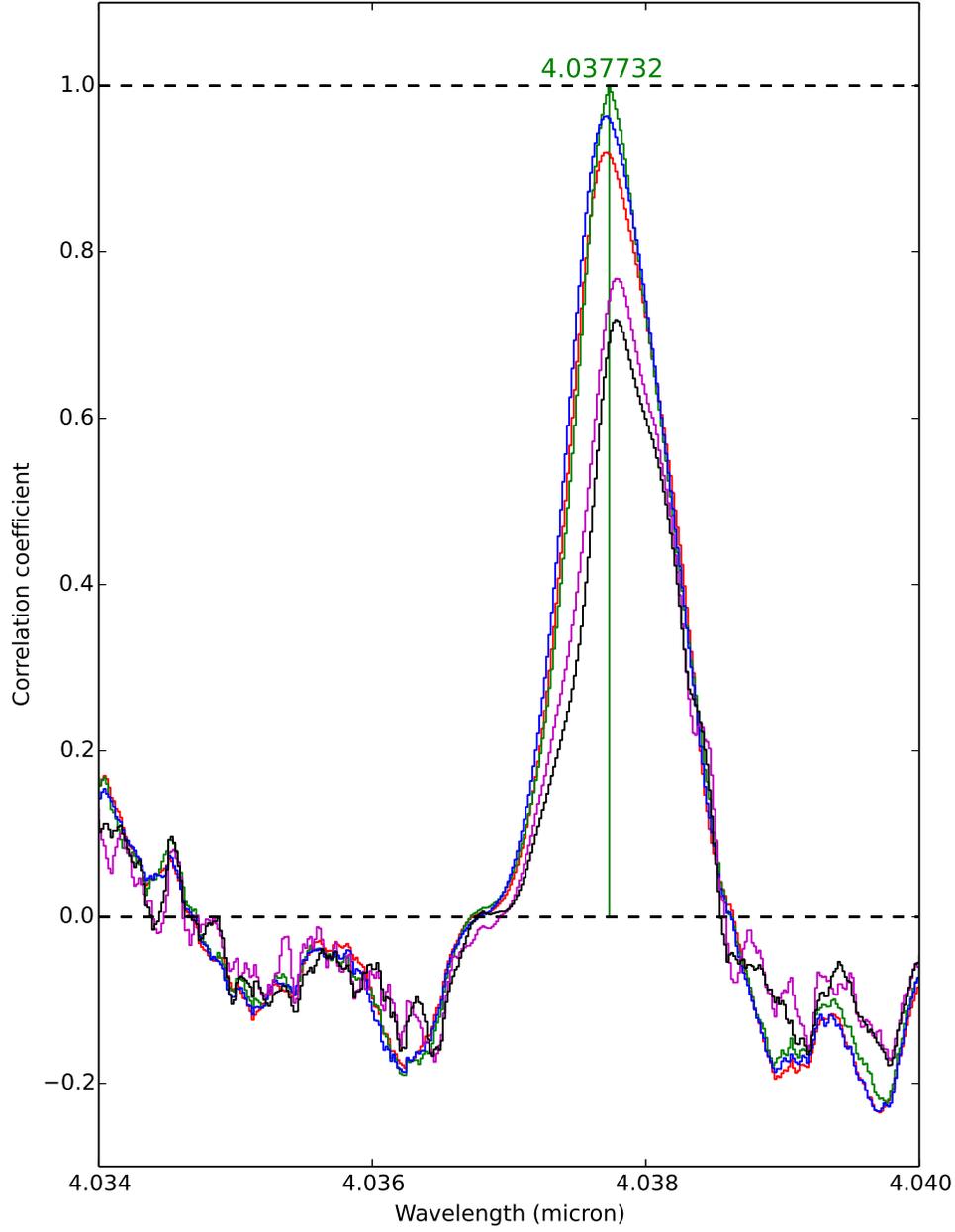}
\caption{Zoomed version of Figure 3, covering the $4.034 - 4.040 \,\mu$m range.
The feature detected here is the He {\small I} $5^3F^0 - 4^3D$ line.} 
\end{figure}
As expected, given the highly ionized nature of this species, its $PV$-diagram is most similar to that of the He\small{ II} 7 -- 6 template.  For clarity, the wavelength region close to the He {\small I} $5^3F^0 - 4^3D$ transition is shown on an expanded scale in Figure 4.  Because this line is itself a template line, the correlation coefficient plotted in green reaches unity at the precise wavelength of the transition.  For the different templates, there are small shifts in the location of the peaks; these reflect small differences in the velocity centroids for the different
template lines.

\begin{deluxetable}{lcrcccccr}
\tabletypesize{\scriptsize}
\tablecaption{Hydrogen recombination lines observed toward NGC 7027}
\tablehead{
iSHELL & Line  & & $\lambda_{\rm obs}$ & $\lambda_{\rm 0}$  & $\lambda_{\rm obs} -  \lambda_{\rm 0}$ & Correl. & Best & Flux$^a$\\
mode  & & & ($\mu$m) & ($\mu$m) &  ($10^{-5} \mu$m)  & coeff. & template & }
\startdata
        L2  &  H {\tiny I}  &     10 -- 5  &   3.039193  &   3.039202  & $-  0.9 \phantom{-}$  &   0.980  &               H {\tiny I}  &  146.3 \cr
       Lp1  &  H {\tiny I}  &      9 -- 5  &   3.296994  &   3.296992  &   0.2  &  0.984  &                H {\tiny I}  &  301.6 \cr
       Lp1  &  H {\tiny I}  &     38 -- 6  &   3.366233  &   3.366265  & $-  3.2 \phantom{-}$  &   0.780  &               H {\tiny I}  &    2.5 \cr
       Lp1  &  H {\tiny I}  &     34 -- 6  &   3.387807  &   3.387845  & $-  3.8 \phantom{-}$  &   0.831  &               H {\tiny I}  &    3.7 \cr
       Lp1  &  H {\tiny I}  &     33 -- 6  &   3.394593  &   3.394558  &   3.5  &  0.809  &              He {\tiny II}  &    6.1 \cr
       Lp1  &  H {\tiny I}  &     32 -- 6  &   3.401878  &   3.401940  & $-  6.2 \phantom{-}$  &   0.880  &               H {\tiny I}  &    4.4 \cr
       Lp1  &  H {\tiny I}  &     31 -- 6  &   3.410077  &   3.410086  & $-  1.0 \phantom{-}$  &   0.868  &               H {\tiny I}  &    4.3 \cr
       Lp1  &  H {\tiny I}  &     30 -- 6  &   3.419025  &   3.419105  & $-  8.0 \phantom{-}$  &   0.941  &               H {\tiny I}  &    6.9 \cr
       Lp1  &  H {\tiny I}  &     27 -- 6  &   3.452850  &   3.452852  & $-  0.2 \phantom{-}$  &   0.789  &               H {\tiny I}  &    7.3 \cr
       Lp1  &  H {\tiny I}  &     26 -- 6  &   3.466959  &   3.466973  & $-  1.4 \phantom{-}$  &   0.903  &               H {\tiny I}  &    8.0 \cr
       Lp1  &  H {\tiny I}  &     25 -- 6  &   3.482948  &   3.482959  & $-  1.1 \phantom{-}$  &   0.929  &               H {\tiny I}  &    8.2 \cr
       Lp1  &  H {\tiny I}  &     24 -- 6  &   3.501149  &   3.501164  & $-  1.5 \phantom{-}$  &   0.946  &               H {\tiny I}  &    9.5 \cr
       Lp1  &  H {\tiny I}  &     23 -- 6  &   3.522025  &   3.522025  & $-  0.0 \phantom{-}$  &   0.974  &               H {\tiny I}  &   11.6 \cr
       Lp1  &  H {\tiny I}  &     22 -- 6  &   3.546100  &   3.546101  & $-  0.1 \phantom{-}$  &   0.966  &               H {\tiny I}  &   13.1 \cr
       Lp1  &  H {\tiny I}  &     21 -- 6  &   3.574152  &   3.574105  &   4.8  &  0.926  &                H {\tiny I}  &   15.3 \cr
       Lp1  &  H {\tiny I}  &     20 -- 6  &   3.606973  &   3.606968  &   0.5  &  0.967  &                H {\tiny I}  &   17.8 \cr
       Lp1  &  H {\tiny I}  &     19 -- 6  &   3.645923  &   3.645923  &   0.0  &  0.972  &                H {\tiny I}  &   21.7 \cr
       Lp2  &  H {\tiny I}  &     18 -- 6  &   3.692635  &   3.692633  &   0.1  &  0.973  &                H {\tiny I}  &   22.2 \cr
       Lp2  &  H {\tiny I}  &      8 -- 5  &   3.740568  &   3.740556  &   1.2  &  0.992  &                H {\tiny I}  &  394.0 \cr
       Lp2  &  H {\tiny I}  &     17 -- 6  &   3.749388  &   3.749393  & $-  0.5 \phantom{-}$  &   0.977  &               H {\tiny I}  &   24.7 \cr
       Lp3  &  H {\tiny I}  &     16 -- 6  &   3.819447  &   3.819451  & $-  0.4 \phantom{-}$  &   0.988  &               H {\tiny I}  &   23.4 \cr
       Lp3  &  H {\tiny I}  &     15 -- 6  &   3.907571  &   3.907549  &   2.2  &  0.974  &                H {\tiny I}  &   34.2 \cr
       Lp3  &  H {\tiny I}  &     14 -- 6  &   4.020869  &   4.020867  &   0.1  &  1.000  &                H {\tiny I}  &   36.3 \cr
       Lp3  &  H {\tiny I}  &      5 -- 4  &   4.052277  &   4.052262  &   1.5  &  0.994  &                H {\tiny I}  &  2470.9 \cr
        M1  &  H {\tiny I}  &     11 -- 6  &   4.672512  &   4.672509  &   0.3  &  0.984  &                H {\tiny I}  &   73.2 \cr

\enddata
\tablenotetext{a}{in a $0.375 \times 15^{\prime\prime}$ slit (units of $\rm 10^{-18}\,W\,m^{-2}$).  The flux measurement uncertainties are dominated by
systematic errors that may be estimated (Section 4.3) as $\pm 20 \%$ (68$\%$ confidence limit).}
\end{deluxetable}

\begin{deluxetable}{lcrcccccr}
\tabletypesize{\scriptsize}
\tablecaption{Helium recombination lines observed toward NGC 7027}
\tablehead{
iSHELL & Line  & & $\lambda_{\rm obs}$ & $\lambda_{\rm 0}$  & $\lambda_{\rm obs} -  \lambda_{\rm 0}$ & Correl. & Best & Flux$^a$\\
mode  & & & ($\mu$m) & ($\mu$m) & ($10^{-5} \mu$m) & coeff. & template &}
\startdata
       Lp2  &  He {\tiny I}  &  $5^3D-4^3P^0$  &   3.703548  &   3.703577  & $-  2.8 \phantom{-}$ & 0.912  &              He {\tiny I}  &    8.0 \cr
       Lp3  &  He {\tiny I}  &  $5^3F^0-4^3D$  &   4.037734  &   4.037734  & $-  0.1 \phantom{-}$ & 1.000  &              He {\tiny I}  &   31.2 \cr
       Lp3  &  He {\tiny I}  &  $5^1F^0-4^1D$  &   4.040941  &   4.040934  &   0.7  & 0.939  &              He {\tiny I}  &    9.2 \cr
       Lp3  &  He {\tiny I}  &  $5^3G-4^3F^0$  &   4.049023  &   4.049013  &   0.9  & 0.990  &              He {\tiny I}  &  110.3 \cr

\\
\hline
\\
        L2  &  He {\tiny II}  &      7 -- 6  &   3.091693  &   3.091693  & $-  0.0 \phantom{-}$  &   1.000  &             He {\tiny II}  &  720.5 \cr
        L2  &  He {\tiny II}  &     14 -- 9  &   3.145466  &   3.145477  & $-  1.1 \phantom{-}$  &   0.992  &             He {\tiny II}  &   29.4 \cr
        L2  &  He {\tiny II}  &    19 -- 10  &   3.151436  &   3.151448  & $-  1.2 \phantom{-}$  &   0.976  &             He {\tiny II}  &    9.4 \cr
       Lp1  &  He {\tiny II}  &    17 -- 10  &   3.484007  &   3.484011  & $-  0.4 \phantom{-}$  &   0.976  &             He {\tiny II}  &   14.0 \cr
       Lp1  &  He {\tiny II}  &    24 -- 11  &   3.490132  &   3.490119  &   1.3  &  0.856  &              He {\tiny II}  &    4.6 \cr
       Lp1  &  He {\tiny II}  &     13 -- 9  &   3.544310  &   3.544315  & $-  0.5 \phantom{-}$  &   0.989  &             He {\tiny II}  &   58.8 \cr
       Lp1  &  He {\tiny II}  &    23 -- 11  &   3.574516  &   3.574577  & $-  6.0 \phantom{-}$  &   0.905  &             He {\tiny II}  &    6.7 \cr
       Lp2  &  He {\tiny II}  &    16 -- 10  &   3.738962  &   3.739029  & $-  6.7 \phantom{-}$  &   0.877  &             He {\tiny II}  &   20.3 \cr
       Lp2  &  He {\tiny II}  &    21 -- 11  &   3.799406  &   3.799424  & $-  1.8 \phantom{-}$  &   0.890  &             He {\tiny II}  &    6.4 \cr
       Lp3  &  He {\tiny II}  &    20 -- 11  &   3.952601  &   3.952618  & $-  1.7 \phantom{-}$  &   0.904  &             He {\tiny II}  &    6.0 \cr
       Lp3  &  He {\tiny II}  &     10 -- 8  &   4.050576  &   4.050609  & $-  3.3 \phantom{-}$  &   0.850  &             He {\tiny II}  &  115.4 \cr
       Lp3  &  He {\tiny II}  &    15 -- 10  &   4.101234  &   4.101248  & $-  1.4 \phantom{-}$  &   0.962  &             He {\tiny II}  &   20.3 \cr
       Lp3  &  He {\tiny II}  &    19 -- 11  &   4.146888  &   4.146914  & $-  2.6 \phantom{-}$  &   0.776  &             He {\tiny II}  &    7.0 \cr
        M1  &  He {\tiny II}  &    17 -- 11  &   4.742625  &   4.742613  &   1.2  &  0.814  &              He {\tiny II}  &    8.6 \cr

\enddata
\tablenotetext{a}{in a $0.375 \times 15^{\prime\prime}$ slit (units of $\rm 10^{-18}\,W\,m^{-2}$).  The flux measurement uncertainties are dominated by
systematic errors that may be estimated (Section 4.3) as $\pm 20 \%$ (68$\%$ confidence limit).}
\end{deluxetable}

This algorithm yields the line lists given in Table 1 for H {\small I} recombination lines and Table 2 for He {\small I} and He {\small II} recombination lines.  To be included in 
these tables, the maximum correlation coefficient had to exceed a threshold that was set at 0.7; with that threshold, every line but one was correctly identified by the algorithm.  That exception was the H {\small I} 33 - 6 transition\footnote{This was a case in which the correlation coefficients with the two template transitions were very nearly equal: 0.809 and 0.804, respectively, for the
He\small{ II} 7 -- 6 and H~\small{I} 14 -- 6 transitions.}, for which the best-correlated template (second column from right) was the He\small{ II} 7 -- 6 transition instead of H\small{ I} 14 -- 6.

Tables 1 and 2 show the observed rest wavelength of each line $\lambda_{obs}$, as determined with the template-fitting algorithm, along with  the laboratory or theoretical determination of the line wavelength, $\lambda_0$.  The values of $\lambda_0$ are those appearing in the Atomic Line List v3.00b4 (hereafter $\rm A^tLL$) compiled by van Hoof (2018).   The rms fractional discrepancy between $\lambda_{obs}$ and $\lambda_{0}$ was $7.4 \times 10^{-6}$
for the lines listed in Tables 1 and 2, corresponding to a Doppler velocity of 
$2.2 \rm \, km\,s^{-1}$ at the wavelengths of relevance.  This is somewhat smaller
than the velocity resolution of the spectrograph ($3.75 \rm \, km\,s^{-1}$).
The maximum fractional discrepancy was $2.3 \times 10^{-5}$.  Although small, these discrepancies greatly exceed the uncertainties in the laboratory or theoretical determinations of the line wavelengths. Thus they provide a measure of the accuracy to which line wavelengths may be determined when the template-fitting algorithm is applied to the NGC 7027 data.  Also shown in Tables 1 and 2 are the peak correlation coefficients achieved at wavelength
$\lambda_{obs}$ and the template that yields the maximum correlation.  The rightmost column
lists the line fluxes, determined by the method that was described in Paper II {and
is discussed briefly below.}

\begin{figure}
\includegraphics[scale=0.7]{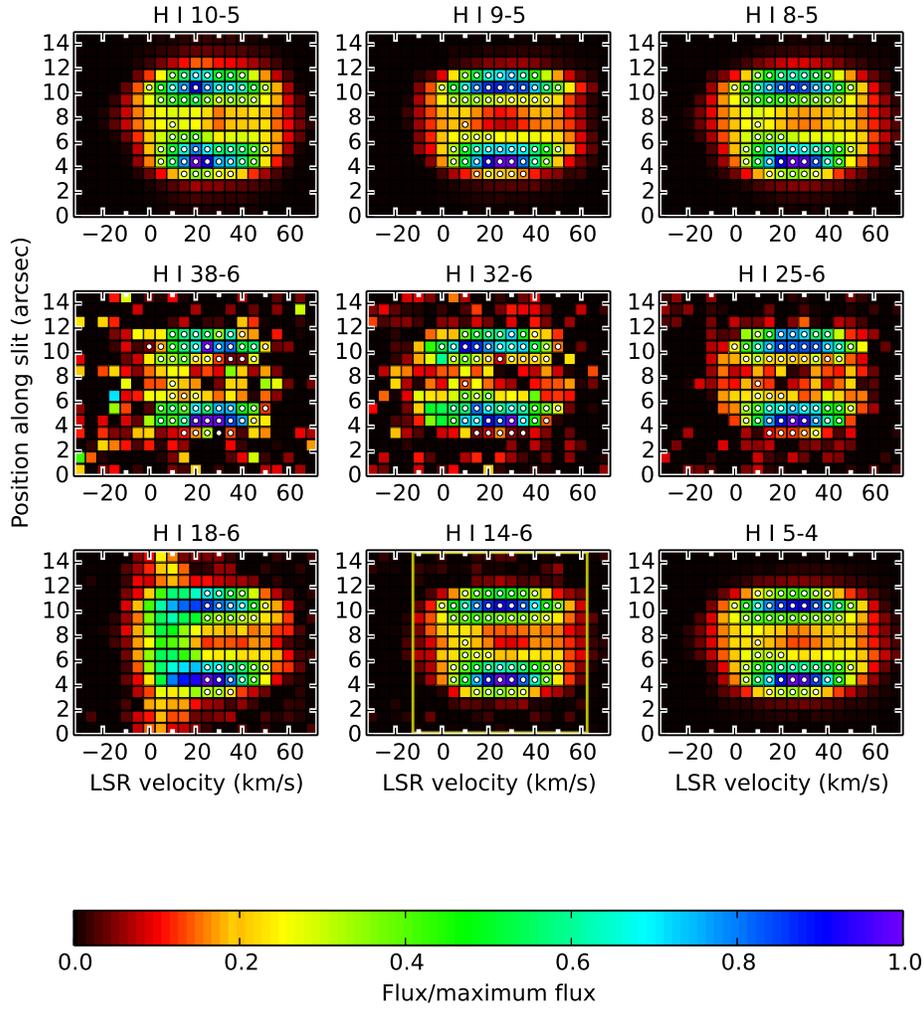}
\caption{{\it PV}-diagrams for a selection of hydrogen recombination lines}
\end{figure}

Figure 5 shows the $PV$-diagrams for a selection of H \small{I} recombination lines.  Each pixel covers 1 arcsec in the spatial dimension and $5 \rm \, km\,s^{-1}$ in the velocity dimension. 
Here, unlike in Figure 2, the intensities are normalized to the maximum value individually in each panel, and the color bar applies to the intensity on a linear, not logarithmic scale.  The yellow rectangle in the panel for H \small{I} $n = 14 - 6$ shows the region 
over which the intensity was integrated to obtain the flux of that line in Table 1.  
White circles indicate the brightest $N_{\rm pix}$ pixels, where $N_{\rm pix}$ 
(47 in this case) is chosen to optimize the signal-to-noise ratio for the total flux in this subset of pixels.  As described in , the fluxes for other H \small{I} lines
are obtained by computing their strength, relative to the H \small{I} $14 - 6$ line,
in this subset of pixels.  This method -- which assumes an identical intensity distribution (in position and velocity) for all H \small{I} lines -- improves the
accuracy of the flux estimates for weak lines.

\begin{figure}
\includegraphics[scale=0.7]{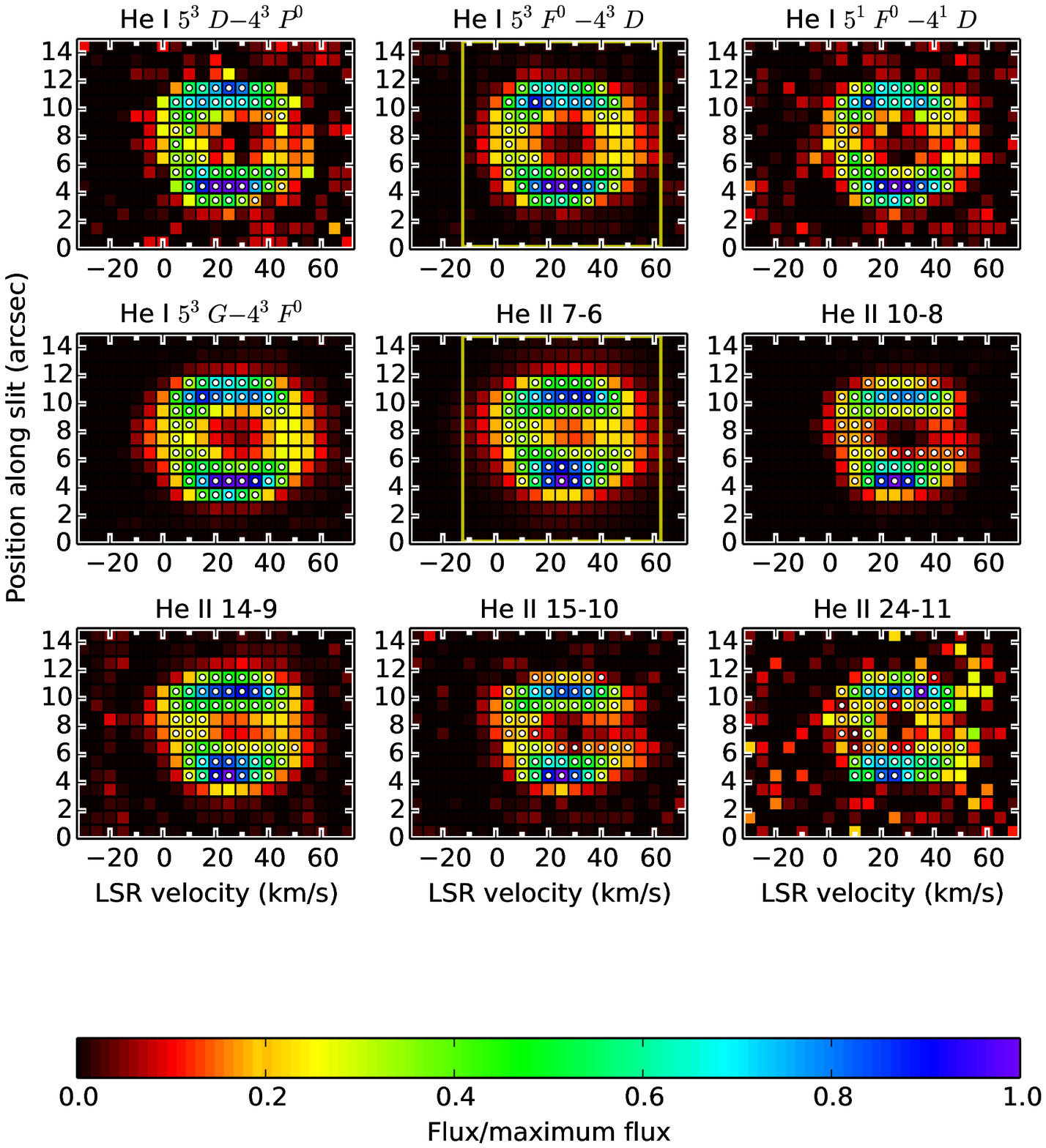}
\caption{{\it PV}-diagrams for a selection of helium recombination lines}
\end{figure}

The H \small{I} recombination lines with $PV$-diagrams shown in Figure 5 have fluxes that
range over three orders of magnitude from the Br$\alpha$ line ($n = 5 - 4$) to the $38 - 6$ line.  The smaller signal-to-noise ratios for the weaker lines are readily apparent
in the plots. For the $18 - 6$ line, a vertical stripe centered near an LSR velocity of $5 \rm \, km\, s^{-1}$ is also apparent.  This arises at the position of a terrestrial atmospheric feature, where the imperfect subtraction of a sky emission line leaves a residual that is independent of slit position.  The affected velocity range was excluded when the flux relative to the $14 - 6$ line was computed, as indicated by the absence of white dots at velocities below $25 \rm \, km\, s^{-1}$.  Figure 6 shows analogous $PV$-diagrams for a selection of He \small{I} and He \small{II} lines.  The relative spatial extents observed for these different sets of recombination lines will be discussed quantitatively at the end of Section 3.

\begin{deluxetable}{lccclrrccr}
\tabletypesize{\scriptsize}
\tablecaption{Metal lines observed toward NGC 7027}
\tablehead{
iSHELL & Line  & & $\lambda_{\rm obs}$ & $\lambda_{\rm 0}$  & $\sigma(\lambda_{\rm 0}) $ & $\lambda_{\rm obs} -  \lambda_{\rm 0}$ &  Correl. & Best & Flux$^a$\\
mode  & & & ($\mu$m) & ($\mu$m) & ($10^{-5} \mu$m) & ($10^{-5} \mu$m) & coeff. & template &}
\startdata
L2 &  [K VII]  	    & $^2P_{3/2}\,-$~$^2P_{1/2}$		& 3.190414	& 3.19053		& 15	&  -12		& 0.887	& He II &  	20.8		\cr
Lp1 & [Fe VII]		& $^3P_{1}\,-$~$^1D_{2}$			& 3.384102	& 3.3846		& 100 	& 50		& 0.657	& He II &  4.8			\cr	
Lp1 & [Zn IV]  	    & $^2D_{3/2}\,-$~$^2D_{5/2}$ 	& 3.624956	& 3.6244		& 120	&  56		& 0.979	& He I  &	45.7		\cr	   
Lp3	& [Ca VII]  	    & $^3P_2\,-$~$^3P_1$  			& 4.086196	& 4.0875		& 150	& -130		& 0.576	& He II &  	4.4		\cr	   
M1 & [Ar VI]  	    & $^2P_{3/2}\,-$~$^2P_{1/2}$		& 4.529302	& 4.5295		& 31	& -20		& 0.501	& He II &  	7170		\cr	    
M1 & [K III]  	    & $^2P_{1/2}\,-$~$^2P_{3/2}$			& 4.617788	& 4.61802		& 47	& -23		& 0.896	& He I &  	126		\cr	    
M1 & [Zn VI]  	    & $^4F_{7/2}\,-$~$^4F_{9/2}$			& 4.689497	& 4.69		  	& 2000	& -50		& 0.965 & He II &  	38		\cr
\\
\hline
\\
L2 & C III      	& 9i -- 8h 					& 3.076921	& 3.0772		& 130	&  -28		& 0.805	& He II &  2.3			\cr	  	  
L2 & C III  	   	& 9k -- 8i						& 3.083853	& 3.0838		& 130	& 5			& 0.919	& He II &  4.6			\cr	   
L2 & C IV	    	& 14f -- 12d 					& 3.086660	& 3.0868		& 210	&  -14		& 0.905	& He I &  5.2			\cr		 
Lp1 & C IV  	    & 11k -- 10i					& 3.281740	& 3.2817		& 150	&  4		& 0.951	& He II & 15.8 			\cr	     
Lp1 & C IV  	   	& 17k -- 14i  					& 3.469021	& 3.4691		& 170	&  -8		& 0.751	& He II &  2.3			\cr	 
Lp3 & C IV  	   	& 15k -- 13i					& 3.867415	& 3.8674		& 210	&  2		& 0.777	& He II &  	3.5		\cr

\\
\enddata
\tablenotetext{a}{in a $0.375 \times 15^{\prime\prime}$ slit (units of $\rm 10^{-18}\,W\,m^{-2}$).  The flux measurement uncertainties are dominated by
systematic errors that may be estimated (Section 4.3) as $\pm 20 \%$ (68$\%$ confidence limit).}
\end{deluxetable}

\begin{figure}
\includegraphics[scale=0.7]{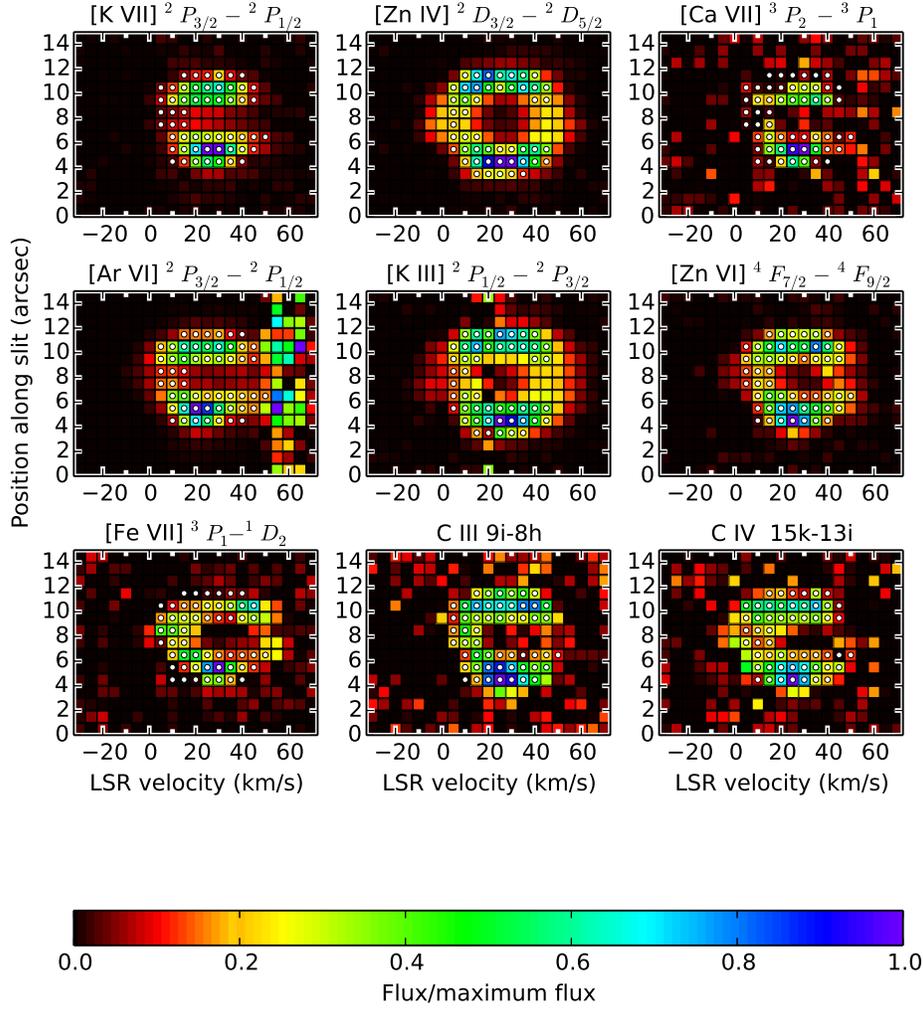}
\caption{{\it PV}-diagrams for a selection of metal lines}
\end{figure}

In Table 3, the fluxes of several metal lines are listed.  These may be divided into
collisionally-excited metal lines (first seven listed, of which six are fine structure transitions within the ground state term) and dipole-allowed lines, populated by recombination (next six lines).  
As in Tables 1 and 2, the rest frequencies of the lines, $\lambda_{\rm obs}$, were measured using the template-fitting method; these are listed, along with the wavelength estimates, $\lambda_0$, appearing in the $\rm A^tLL$ (van Hoof 2018).  In the case of these metal lines, the previous wavelength estimates are typically far more uncertain than the current measurements.  In addition to showing the difference $\lambda_{\rm obs} - \lambda_0$, I have therefore also shown the stated uncertainties for $\lambda_0$ (here denoted $\sigma(\lambda_0)$ and obtained from the $\rm A^tLL$).   In all cases, $\vert \lambda_{\rm obs} - \lambda_0 \vert \le \sigma(\lambda_0)$, providing support for the line identification.  Figure 7 shows the $PV$-diagrams for a selection of the metal lines listed in Table 3.

Of the six fine-structure lines listed in Table 3, all except the [Zn \small{VI}] line have been identified previously in planetary nebulae (e.g. Beintema \& Pottasch 1999; Dinerstein \& Geballe 2001).  Indeed, for the [K \small{VII}], [Ar \small{VI}], and  
[K \small{III}] transitions the recommended 
values of $\lambda_0$ appearing in the $\rm A^tLL$ are astronomical determinations (Feuchtgruber et al.\ 1997).  As far as I aware, 
the line near 4.689497~$\mu$m has not been reported previously.  My proposed
identification of this line as the $^4F_{7/2} -$~$^4F_{9/2}$ transition within the ground state term of Zn$^{5+}$ will be discussed below in Section 4.2.

Another spectral feature, near 3.384102~$\mu$m, can be identified as the 
$^3P_{1} - $~$^1D_{2}$ transition within the [Ar]~$3d^2$ ground-state configuration
of [Fe~VII].  This identification is supported by a previous detection 
(Zhang et al.\ 2005) of the 4894$\AA$ $^3P_{1} - $~$^3F_{2}$ transition.
This optical wavelength transition, which originates in the same upper state, had a 
source-integrated (and extinction-corrected)
flux of $6.2 \times 10^{-4}$ relative to the H$\beta$ line.  Given the
relative photon energies and spontaneous radiative decay rates for the two [Fe VII] transitions, and given the hydrogen line intensities expected for Case B recombination,
this value would imply a [Fe VII] 3.384102~$\mu$m to Br$\alpha$ ratio of $1.47 \times 10^{-3}$.  The observed value (from Tables 1 and 3) is $1.88 \times 10^{-3},$ in reasonable agreement with the expected value.\footnote{Exact agreement is not expected anyway, since the flux measurements reported here were obtained with a narrow slit positioned along the minor axis of the nebula; those given by Zhang et al.\ (2005), by contrast, were integrated over the
entire source. Because the [Fe VII] emission is more compact than the H~\small{I} emission, the measured {[Fe VII]} 3.384102~$\mu$m to Br$\alpha$ ratio will be enhanced relative to the source-integrated value.}

Six additional spectral lines are plausibly identified as C \small{III} and C \small{IV} recombination lines.  For the C \small{III} transitions, 
the configurations of the upper and lower states are $1s^2 2s nl$ and thus the total spins are 0 or 1.  However, because the energy differences between the singlet and triplet states are very small, and because only one electron has non-zero orbital angular momentum, specifying $nl$ suffices to identify the upper and lower states.
Recombination lines of C \small{III} and C~\small{IV} have been previously detected toward NGC 7027 at optical wavelengths (Baluteau et al.\ 1995; Zhang et al.\ 2005).

Table 3, like Tables 1 and 2, indicates which of the five line templates yields the best correlation with the $PV$-diagram.   {For three of the fine-structure transitions, the maximum correlation coefficient did not exceed the threshold of 0.7 that I adopted in
compiling the list of recombination lines.  
These three lines comprise the weakest two lines, [Fe~VII] and [Ca~VII], and the [Ar~VI] line that is very strong but suffers interference from a nearby telluric feature.}

As expected, the more highly ionized species typically resemble the He \small{II} template more than any other template, while the lower ionization species (Zn$^{3+}$ and K$^{2+}$) resemble He \small{I} most closely.
Spatial moments provide an alternative method of quantifying the distribution of the observed emission for each spectral line.  In particular, we may consider the quantity

\begin{equation}
\theta_{\rm rms} = \sqrt {\int I(\theta) (\theta-\theta_c)^2 d\theta \bigg/ \int I(\theta)  d\theta}
\end{equation}
where $I(\theta)$ is the velocity-integrated intensity at angular position $\theta$
along the slit, and $\theta_c$ is the centroid position:
\begin{equation}
\theta_c = \int I(\theta) \theta d\theta 
\end{equation}

\begin{figure}
\includegraphics[scale=0.7]{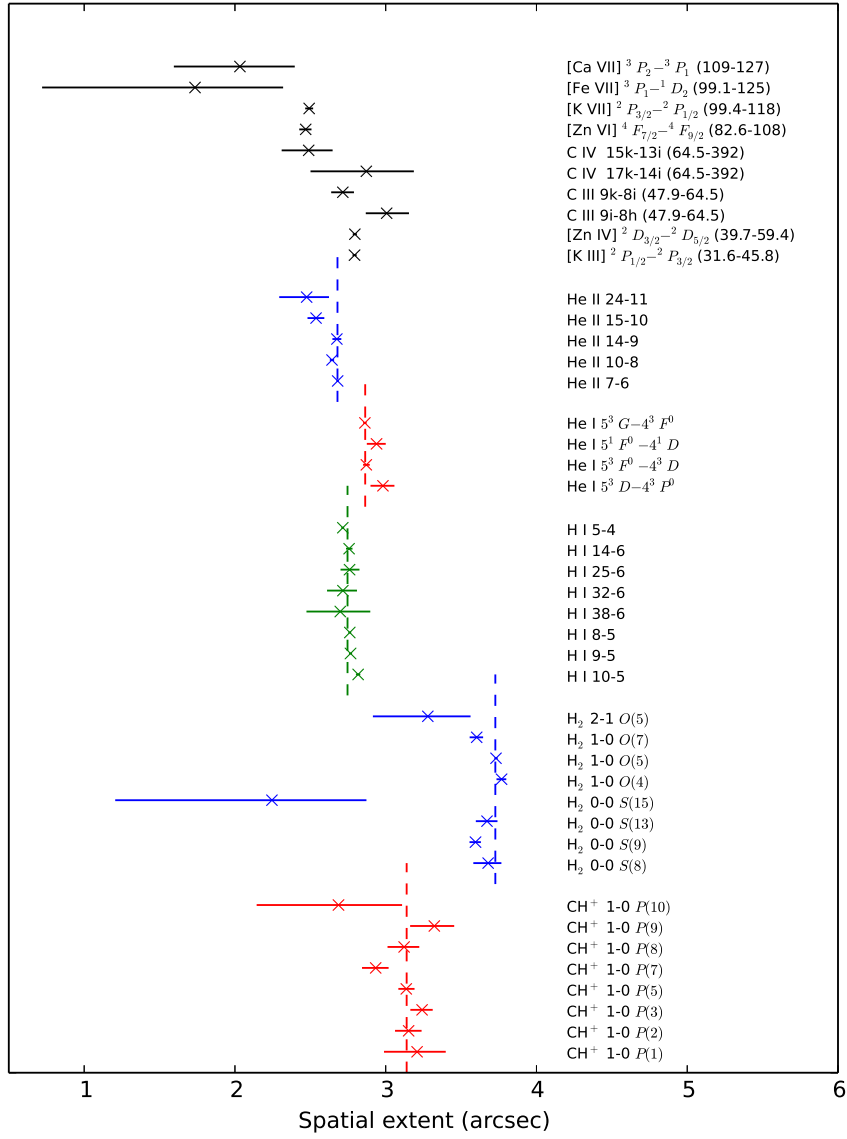}
\caption{Spatial extent of selected lines, $\theta_{\rm rms}$ (see the text for the definition)}
\end{figure}

The quantity $\theta_{\rm rms}$, which represents the rms angular displacement of the observed emission from its centroid, is plotted in Figure 8 for
a selection of spectral lines (which include the CH$^+$ and H$_2$ emissions reported
in Paper II).  The error bars shown here represent $68.3\%$ confidence limits.  These were obtained by computing $\theta_{\rm rms}$ in a set of 1000 simulations in which Gaussian random noise was added to the data with an amplitude equal to the noise observed in line-free 
spectral regions.  Vertical dashed lines show the weighted means of $\theta_{\rm rms}$ for lines of a given species.  

The results shown in Figure 8 indicate that the molecular emissions are significantly more
extended than the emissions from atoms and atomic ions, as expected because the molecular
gas lies outside the photoionized nebula.  The He \small{II} lines have a smaller $\theta_{\rm rms}$ than the
H \small{I} lines, consistent with expectation that the He$^{2+}$ zone is smaller than the
H$^+$ zone.  On the other hand, the He \small{I} lines have a larger $\theta_{\rm rms}$ than the H \small{I} lines.  This behavior is also expected: while the outer edge of the He$^+$ zone is roughly cospatial with that of the H$^+$ zone, the He$^+$ zone also has an inner edge within which He$^+$ is absent (having been ionized to form He$^{2+}$).

For the metal lines in Figure 8, shown in black, two key energies are listed in parentheses with the notation $I_1 - I_2$.  Here, $I_1 \rm \, eV$ is the appearance potential (i.e.\ the minimum photon energy needed to produce the relevant ion by photoionization); and $I_2 \rm \, eV$ is the ionization potential for that ion.  In this context, the ``relevant ion" is the emitting ion for the collisional excited lines (e.g.\ K$^{2+}$ for [K \small{III}], Ar$^{5+}$ for [Ar \small{VI}]).  For recombination lines, however, the relevant ion is the recombining ion (e.g.\ H$^+$ for HI, C$^{4+}$ for C \small{IV}).  The results shown in Figure 8 show the expected trend that higher ionization species show emission with a smaller spatial extent. 

\section{Discussion}

\subsection{Improved wavelength determinations for infrared metal lines}

The wavelengths of many infrared fine-structure transitions are poorly known from 
laboratory experiment.  In many cases, they must be estimated by analyzing differences in
the measured wavelengths of dipole-allowed ultraviolet transitions at much
higher frequencies.  The fractional accuracy of the derived infrared 
wavelengths can therefore be worse than that of the measured ultraviolet wavelengths
by two orders of magnitude or more. In the case of [Zn \small{VI}] $^4F_{7/2} - $~$^4F_{9/2}$ for example, which has the wavelength that is least well known, the estimate of $\lambda_0$ in Table 3 was derived  from spectroscopic measurements of the $3d^6 4p - 3d^7$ system in the $209 - 284 \AA$ region (van~Het~Hof et al.\ 1994).   The wavelength accuracy was $0.004 \AA$, which is equivalent to a wavenumber accuracy of 5 to $9 \, \rm cm^{-1}.$  At the frequency of the $^4F_{7/2} -$~$^4F_{9/2}$ transition ($\sim 2132 \,\rm cm^{-1}$), this corresponds to a fractional accuracy of 0.2 -- 0.4$\%$).  
Theoretical quantal calculations of spin-orbit coupling generally fare no better, with state-of-the-art techniques providing typical accuracies no better than 
$\sim 10\, \rm cm^{-1}$ (Lan Cheng, personal communication).

For fine-structure transitions that are observed from gaseous nebulae, astronomical
observations sometimes provide the best wavelengths currently available.  Such is the case for the [K \small{VII}], [Ar \small{VI}], and [K \small{III}] transitions listed in Table 3, for which the wavelength estimates $\lambda_0$ were obtained from Infrared Space Observatory (ISO) observations of the planetary nebulae NGC 7027 and NGC 6302 (Feuchtgruber et al.\ 1997).  These observations were performed with the SWS grating spectrometer which had a spectral resolving power $R = \lambda / \Delta \lambda \sim 1000 - 2000.$  The observations reported here, performed at considerable higher resolution ($R = 80,000$) with the iSHELL spectrometer, yield substantial accuracy improvements for the three transitions that were considered by Feuchtgruber et al.\ (1997); for the other transitions, the gains in accuracy are even larger.  The values of $\lambda_{\rm obs}$ given in Table 3 are
therefore the most accurate values obtained to date.  The integrated spectra for a selection of metal ions are shown in Figure 9.

\begin{figure}
\includegraphics[scale=0.7]{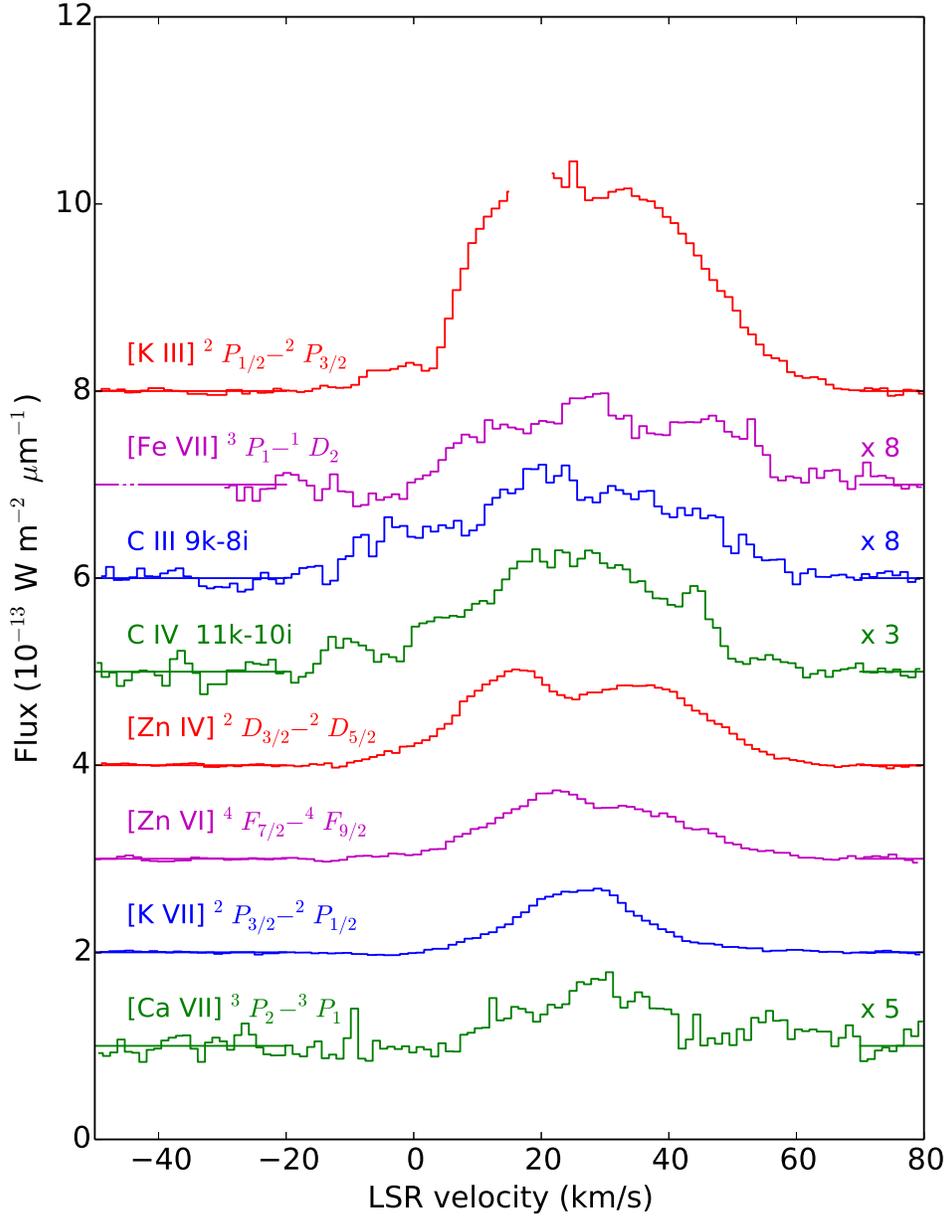}
\caption{Spectra of selected metal lines, obtained for the full slit}
\end{figure}

There are two sources of uncertainty in the values of  $\lambda_{\rm obs}$ that are obtained.  The first relates to the fundamental accuracy of the wavelength measurements obtained from the new observational data and may be evaluated by considering the results shown in Table 1 and 2.  Because these involve lines for which the wavelengths $\lambda_0$ have negligible uncertainties, we may adopt the rms value of $(\lambda_{\rm obs}-\lambda_0)/\lambda_0$, or $7.4 \times 10^{-6}$, as an estimate of the fractional accuracy of the wavelength measurements.  The second source of uncertainty arises from the possibility that the metal line emissions 
may have a systematic Doppler shifts relative to the best-fit templates (He \small{I} or 
He \small{II}).  This possibility is suggested by a careful examination of Figure 4, in which the five curves peak at slightly different wavelengths.  Considering only the templates for atomic recombination lines, the maximum fractional variation is $6 \times 10^{-6}$.  The quadrature sum of two fractional uncertainties given above is $1 \times 10^{-5}$, which may be taken as a reasonable estimate of the fractional accuracy of the improved metal line wavelengths reported here.  This represents a factor 5 -- 15 improvement over the previous astronomical wavelength estimates, and a factor 30 - 400 improvement over the laboratory measurements.

\subsection{Probable detection of the [Zn~VI] $^4F_{7/2},-$~$^4F_{9/2}$ fine structure  transition}

A spectral line is detected with high confidence with a rest wavelength of 
$4.689497 \pm 0.000047\, \mu \rm m$: the $PV$-diagram is shown in Figure 6 
(middle row, right column). 
Its probable identity is the $^4F_{7/2}\, - $~$^4F_{9/2}$ fine structure transition 
within the (inverted) ground state $^4F$ term of [Zn~VI], which has the configuration  [Ar]$\,3d^7$.  
This identification is supported by a number of arguments: (1) this is the only plausible identification among the lines catalogued in the $\rm A^tLL$; (2) the spatial extent, $\theta_{\rm rms}$, of the $4.6895\, \mu \rm m$ line (Figure 8) is in excellent agreement with that of the $^2P_{3/2}\, - $~$^2P_{1/2}$ line of [K~VII], which is emitted by an ion with ionization and appearance potentials similar to those of Zn$^{5+}$; (3) another zinc ion has previously been identified in NGC~7027 by Dinerstein \& Geballe (2001) (and detected again in the observations reported here): the Zn$^{3+}$ ion, with its $^2D_{3/2}\,-$~$^2D_{5/2}$ transition near $3.62496\, \mu \rm m$.
Although there is no evidence for the higher-lying [Zn~VI] $^4F_{5/2}\, - $~$^4F_{7/2}$ transition near 4.03~$\mu$m, electron impact excitation from the $^4F_{9/2}$ ground state to the upper $^4F_{5/2}$ state for that transition is probably disfavored because collision strengths typically decline with $\Delta J$.

\begin{figure}
\includegraphics[scale=0.7]{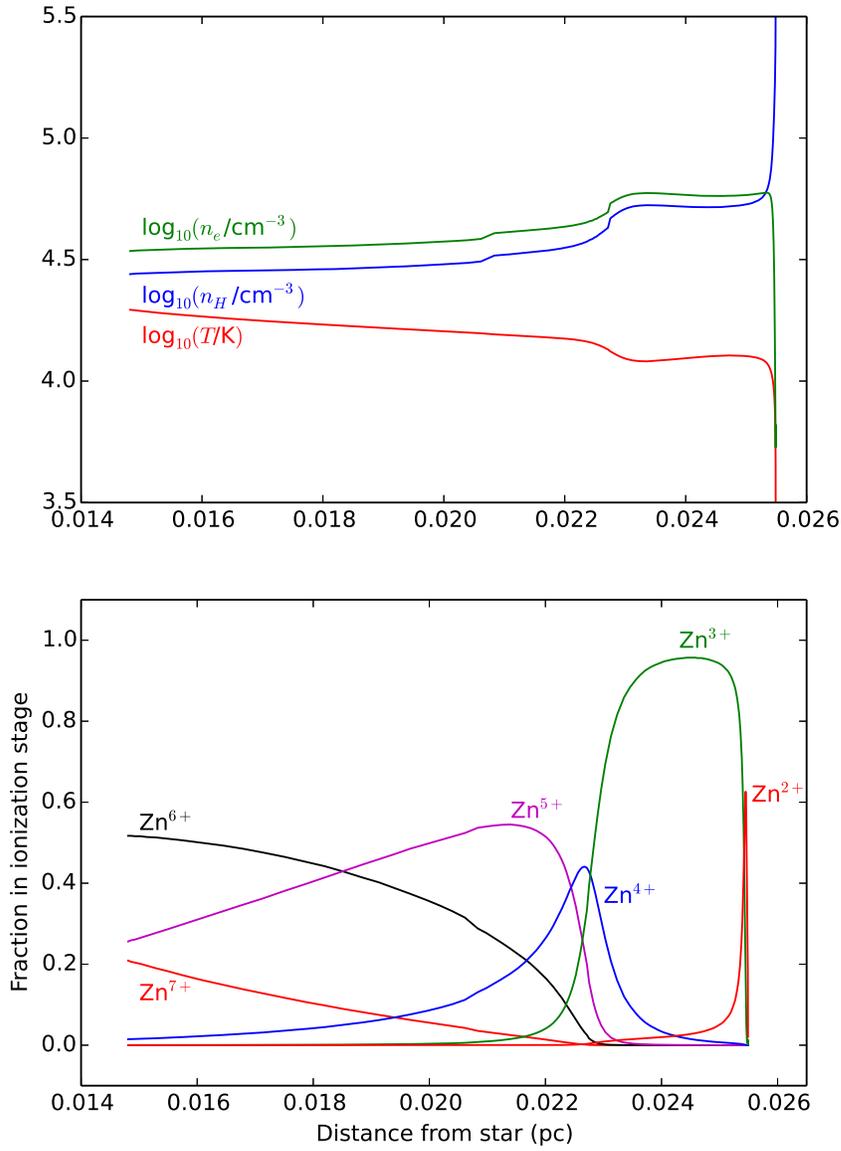}
\caption{CLOUDY model predictions for zinc ions}
\end{figure}

To further evaluate the plausibility of this identification, I have obtained predictions
for the ionization state of Zn using the CLOUDY photoionization model v17.02
(Ferland et al.\ 2017).   Full details of the model parameters have been described in Paper I, where CLOUDY was used
to model the HeH$^+$ cation: the central star, with a photospheric temperature of $1.9 \times 10^5$~K is assumed to irradiate a constant pressure shell.  The lower panel in Figure 10 shows how the ionization state of zinc varies with distance from the central star, for an adopted distance of 980~pc (see Paper I); the upper panel shows the variation
of the electron temperature, $T$, the electron density, $n_e$, and the density of 
H nuclei, $n_{\rm H}$.

Based upon the ionization structure shown in Figure 10, I have computed the line emissivities for the [Zn~IV] and [Zn~VI] transitions listed in Table 3.  As Smith et al. (2014) pointed out for the case of the [Zn~IV] transition, the estimated density of the nebula lies comfortably below the critical density above which collisional deexcitation becomes dominant.  Thus, almost all Zn$^{3+}$ ions will be in the ground $^2D_{5/2}$ fine structure state, and every excitation will be followed by the emission of a $3.62496\, \mu \rm m$ photon.  Because the radiative decay rates for the [Zn~VI] and [Zn~IV] transitions are very similar, the same argument will apply to the [Zn~VI] transition.  With these assumptions, the emissivities are simply proportional to the elemental abundance ratio, $n_{\rm Zn}/n_{\rm H}$, and the dimensionless collision strength, $\Upsilon$.  

Integrating the emissivity over the region enclosed by the iSHELL slit, the model yields a flux of $2.5 \times 10^{-17} \Upsilon_{3.6250} (10^8\,n_{\rm Zn}/n_{\rm H}) \rm \,W \,m^{-2}$ for the [Zn~IV] $^2D_{3/2}\, -$~$^2D_{5/2}$ transition.  Given the observed flux listed in Table 3 and an estimated collision strength $\Upsilon_{3.6250}=1.4$ (Smith et al. 2014, citing a private communication by Butler), the zinc abundance is found to be $1.3 \times 10^{-8}$, in perfect agreement with the estimate of Smith et al.\ (2014).  This amounts to roughly one-third the solar abundance of Zn (Asplund et al.\ 2005).

For the [Zn~VI] $^4F_{7/2}\, - $~$^4F_{9/2}$ transition, the predicted flux is $4.9 \times 10^{-18} \Upsilon_{4.6895} (10^8\,n_{\rm Zn}/n_{\rm H})\rm \,W \,m^{-2}$, which would require $\Upsilon_{4.6895} (10^8\,n_{\rm Zn}/n_{\rm H}) = 7.8$.  
I am not aware of any 
theoretical estimates for the collision strength $\Upsilon_{4.6895}$; given the zinc abundance determined above from the [Zn~IV] transition, the required value is 6.0.  This value is fairly large, but by no means outside the typical range (1 to 10; e.g.\ Draine 2011).  To summarize, although my analysis of the $4.6895\, \mu \rm m$ line flux is limited by absence of available collision strengths, it does indicate that the [Zn~VI] $^4F_{7/2}\, - $~$^4F_{9/2}$ transition can very plausibly account for the observed flux of the $4.6895\, \mu \rm m$ line 

\subsection{Relative line fluxes for recombination lines of H and He$^+$}


The availability of reliable flux measurements for 39 infrared recombination lines of 
H~{\small I} and He~\small{II} 
enables a test of model predictions for their relative strengths.   
In Figure 11, the H~\small{I} line fluxes are shown relative to the H~\small{I} $n=14 - 6$ line, on a logarithmic scale and as a function of the principal quantum number of
the upper state, $n_U$.  As indicated in the legend in the upper right, different colors
denote different principal quantum numbers for the lower state, $n_L$; the H~\small{I} lines observed in this spectral region include one Brackett line (Br$\alpha$, with $n_L=4$), two lines from the Pfund series ($n_L=5$), and 22 lines from the Humphreys series ($n_L=6$, with $n_U$ covering the wide range from 10 to 38).  Solid lines show the model predictions presented by Storey \& Hummer (1995, hereafter SH95), for an assumed electron temperature of $1.5 \times 10^{4}$~K and electron density of $10^{5} \rm\,cm^{-3}$.   The predictions were obtained for ``Case B" recombination
(i.e.\ under the assumption that the Lyman lines are optically-thick).   

\begin{figure}
\includegraphics[scale=0.7]{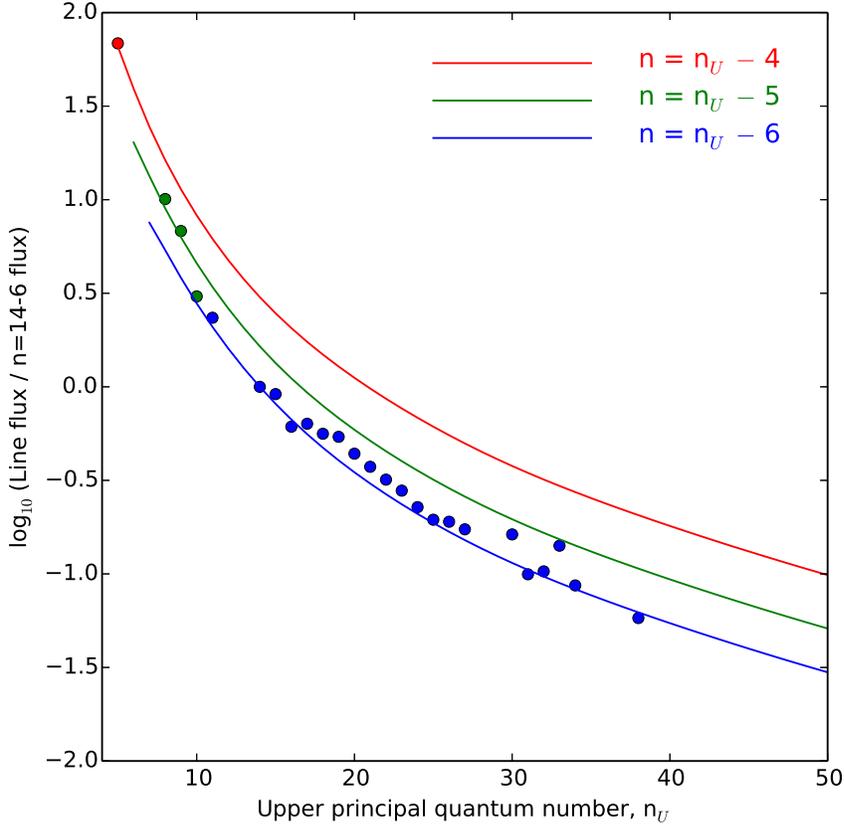}
\caption{H~I recombination line ratios.  { The line fluxes are presented as a 
ratio to the H 14 -- 6 line, as a function of the principle quantum number of the 
upper state, $n_U$.  Solid curves show the theoretical predictions (SH95) for ``Case B" recombination with an assumed electron temperature of $1.5 \times 10^{4}$~K and electron density of $10^{5} \rm\,cm^{-3}$.  The observational data points and theoretical predictions are color-coded according the the principle quantum number of the lower state
(see inset legend).}}
\end{figure}

The rms deviation of $\rm log_{10}$(Line~flux~/~$n=14-6$~flux) from the model prediction is 0.078, corresponding to an fractional rms deviation of $20 \%$.  This then provides a 
validation of the SH95 model predictions over a wide range of $n_U$.
Barring some conspiracy between the flux measurement errors and errors in the model predictions, it suggests a limit of $20 \%$ on the typical systematic errors for flux ratio measurements.  It also confirms that the effects of differential dust extinction
are negligible over the 3.0 - 4.7~$\mu$m range, as expected\footnote{Given a reddening of 3.4 mag inferred by Zhang et al.\ (2005) at the H$\beta$ wavelength and the infrared 
extinction curve presented by Mathis (1990; his Table 1), the extinction is expected 
to be only 0.15 mag at 3.4~$\mu$m and 0.08 mag at 5~$\mu$m.}.

\begin{figure}
\includegraphics[scale=0.7]{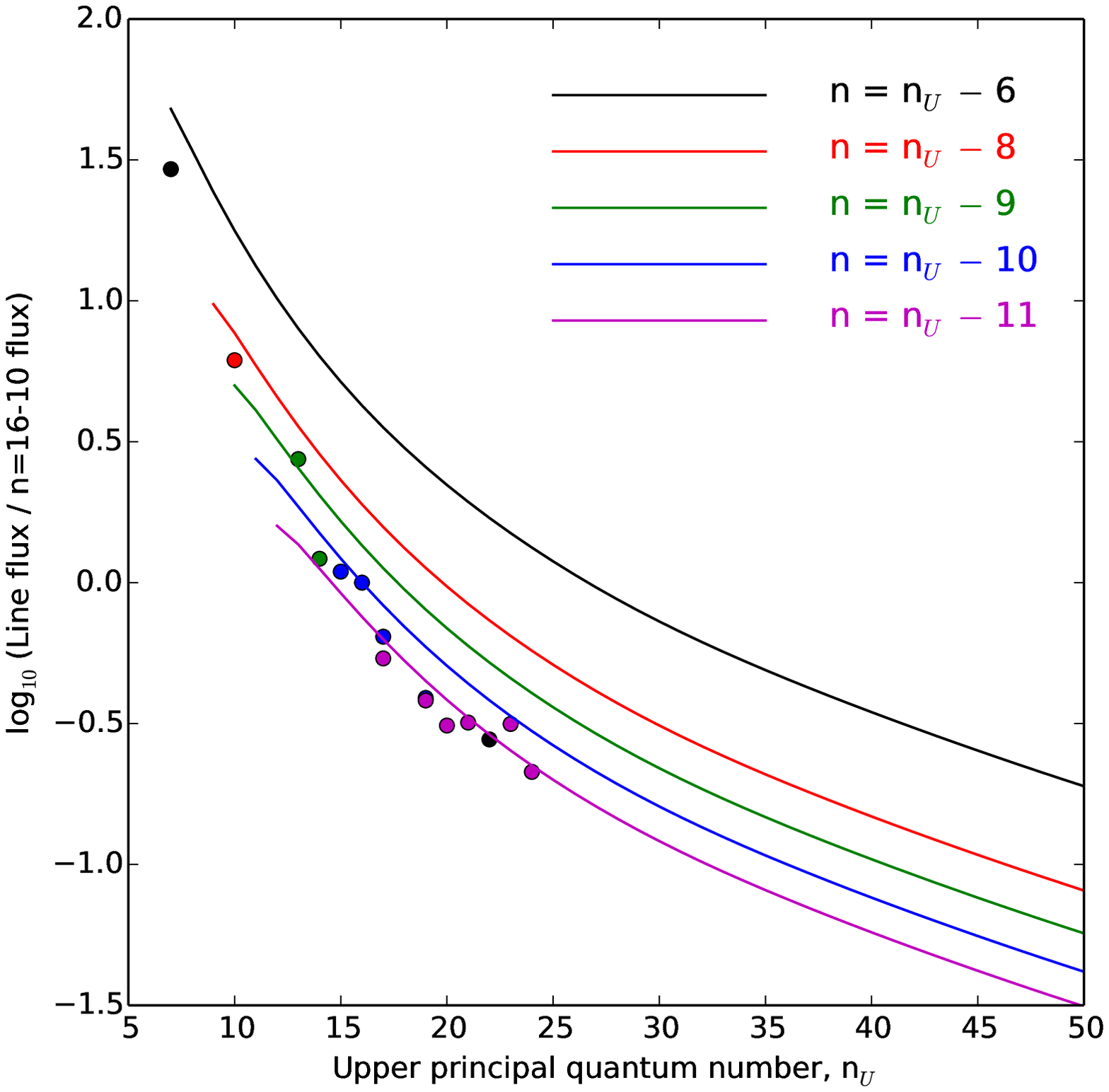}
\caption{Goodness of fit for H~I recombination line ratios in the
$T_e-n_e$ plane.  Red, green and blue contours show the 68$\%$, 95$\%$ and 99.7$\%$
confidence limits.  The black locus shows the run of $n_e$ with $T_e$ in the ionized
gas (from the CLOUDY model prediction shown in Figure 10, upper panel).}
\end{figure}

\begin{figure}
\includegraphics[scale=0.7]{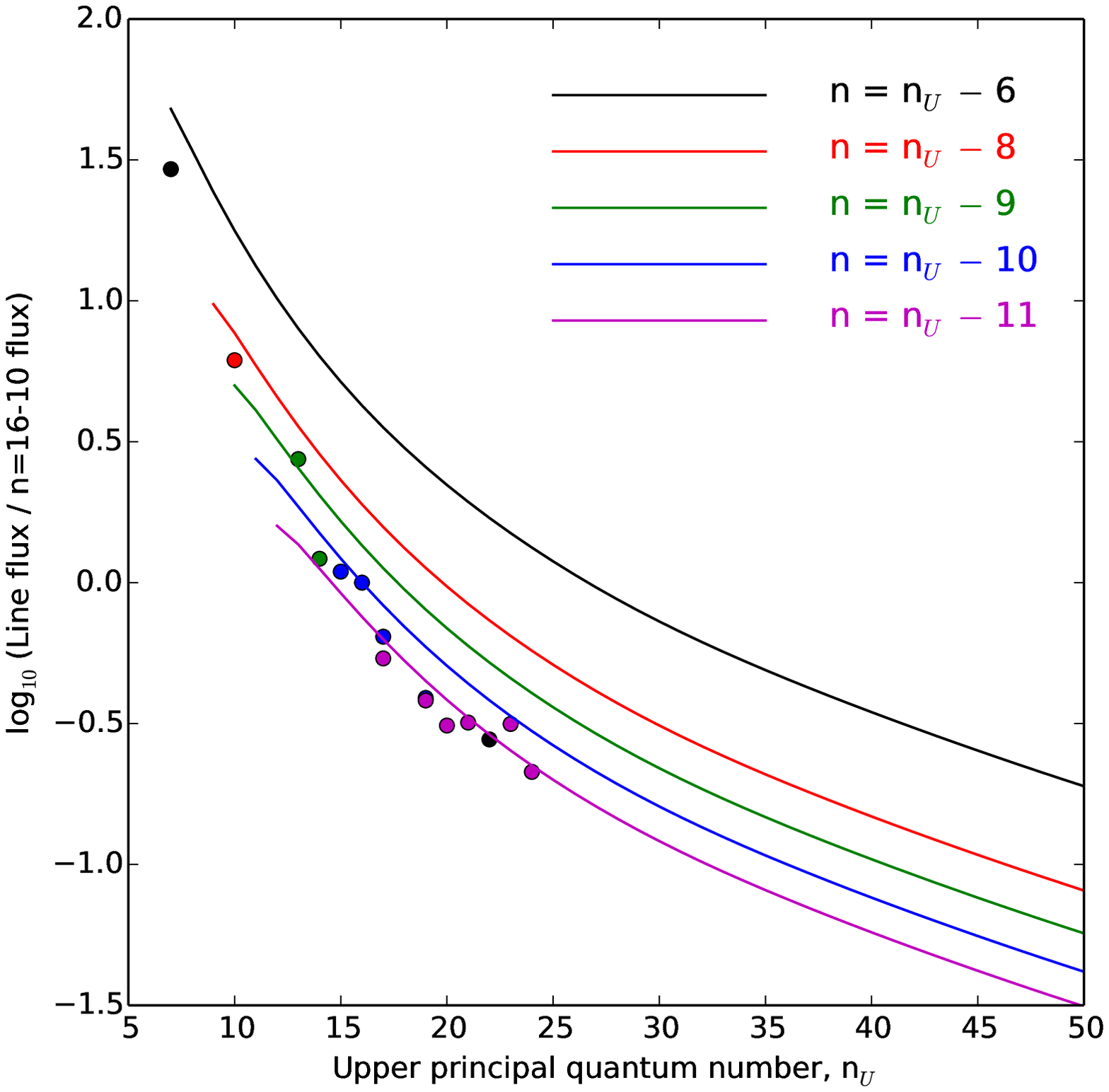}
\caption{He~II recombination line ratios.  {The line fluxes are presented as a 
ratio to the He~II 16 -- 10 line, as a function of the principle quantum number of the 
upper state, $n_U$.  Solid curves show the theoretical predictions (SH95) for ``Case B" recombination with an assumed electron temperature of $1.5 \times 10^{4}$~K and electron density of $10^{5} \rm\,cm^{-3}$.  The observational data points and theoretical predictions are color-coded according the the principle quantum number of the lower state
(see inset legend).}}
\end{figure}

The SH95 model predictions have been obtained over a range of electron temperatures $T$ from 
3000 to 20,000~K and a range of electron densities $n_e$ from 10$^2$ to 10$^{14}\,\rm cm^{-3}$.  Figure 12 shows the goodness of fit in the  $T-n_e$ plane, for models with a constant temperature and density; 
here red, green and blue contours represent the 
68, 95, and 99.7$\%$ confidence limits.
The black line indicates the locus of temperature and density that is predicted by the CLOUDY model (Section 4.2) within the region where hydrogen is mainly ionized.   Because the predicted line ratios are only weakly dependent on $T$ and $n_e$, the recombination line ratios are not strongly constraining: only unrealistically high or low electron densities -- or unrealistically low electron temperatures -- are excluded at the 95$\%$ confidence level.  Figure 13 shows the line fluxes observed for recombination lines of He~\small{II}, compared with the corresponding SH95 predictions.
Here the measured line fluxes are shown relative to the $n=16-10$ line flux, and the 
rms deviation of $\rm log_{10}$(Line~flux~/~$n=16-10$~flux) from the model prediction is 
0.116.  The set of observed lines has $n_U$ ranging from 7 to 23 and $n_L$ ranging from 6 to 11.

\section{Summary}

\noindent 1) An infrared L- and M-band spectral survey has been performed toward the young planetary nebula NGC 7027 with the iSHELL instrument on NASA's Infrared Telescope Facility (IRTF).
It covers the 2.951 - 5.24 micron spectral region (incompletely) at a spectral resolving power of 80,000 and provides spatial information along a 15 arcsec slit.

\noindent 2) When the spectral and spatial information from this survey is analyzed with the template-fitting algorithm described in Section 3, 56 transitions of atoms or atomic ions are detected with high confidence.

\noindent 3) A spectral line at 4.6895$\rm \,\mu$m is most probably the $^4F_{7/2}\, - $~$^4F_{9/2}$ fine structure transition of Zn$^{5+}$, an ion that has not previously been detected in gaseous nebulae.  An ionization model for NGC 7027, obtained with the 
CLOUDY photoionization code, predicts Zn$^{5+}$ to be the dominant ionization stage of zinc in the middle part of the irradiated shell.  This model can
simultaneously fit the 4.6895$\rm \,\mu$m line together with a [Zn IV] fine-structure line that has also been detected, given an assumed collisional strength of 6 for the 
electron impact excitation of Zn$^{5+}$ from the ground $^4F_{9/2}$ state to the $^4F_{7/2}$ state.

\noindent 4) The rest wavelengths of 13 metal lines - including six infrared fine structure transitions - are determined to an unprecedented fractional accuracy of $1 \times 10^{-5}$.  They are presented in Table 3.

\noindent 5) The relative strengths of 39 recombination lines of H and He$^+$, with upper states of principal quantum number up to 38 (H) or 24 (He$^+$), agree well with
predictions obtained for Case B recombination (SH95).

\begin{acknowledgements}

The observations reported here were carried out at the Infrared Telescope Facility (IRTF), 
which is operated by the University of Hawaii under contract NNH14CK55B with the 
National Aeronautics and Space Administration.  I am very grateful to the IRTF director, John Rayner, for making unallocated engineering time available for the July 2019 observations that initiated this project; and to the IRTF support astronomers and telescope operators, for the excellent support they provided for the
observations reported here.  I thank Rolf G\"usten for many helpful suggestions that improved a previous version of the manuscript, {and the anonymous referee for
several useful comments}.
Finally, I am particularly grateful to Miwa Goto, who expertly led the acquisition of the data at the IRTF.

\end{acknowledgements}

\end{document}